\documentclass[12pt]{article}
\usepackage{amsmath,amssymb}
\usepackage[dvips]{epsfig}
\newtheorem{theorem}{Theorem}[section]
\newtheorem{lemma}[theorem]{Lemma}

\newtheorem{proposition}[theorem]{Proposition}

\newtheorem{remark}[theorem]{Remark}
\protect
\newcommand{\res[1]}{\operatornamewithlimits{Res}_{#1}}
\newcommand{\vu}{\boldsymbol{u}}

\newcommand{\lb}{\lambda}
\newcommand{\vb}{\boldsymbol{\beta}}
\begin{document}

\title{Modulation of Camassa--Holm equation and reciprocal transformations}

\author {Simonetta Abenda \\
Dipartimento di M
atematica e CIRAM \\
Universit\`a degli Studi di Bologna, Italy \\
{\footnotesize  abenda@ciram.unibo.it} \\
and \\
Tamara Grava \\
SISSA, Via Beirut 9, Trieste, Italy  \\
{\footnotesize  grava@sissa.it}} \maketitle
\begin{center}
{\it Dedicated to Pierre van Moerbeke on his sixtieth birthday}
\end{center}
\begin{abstract}
We derive the modulation equations or Whitham equations for the
Camassa--Holm (CH) equation. We show that the modulation equations
are hyperbolic and  admit bi-Hamiltonian structure. Furthermore
they  are connected  by a reciprocal transformation to the
modulation equations of the first negative flow of the Korteweg de
Vries (KdV) equation. The reciprocal transformation is generated
by the Casimir of the second Poisson bracket of the KdV averaged
flow. We show that the geometry of the bi-Hamiltonian structure of
the KdV and CH modulation equations is quite different: indeed the
KdV  averaged  bi-Hamiltonian structure can always be related to a
semisimple Frobenius manifold while the CH one cannot.
\end{abstract}

\section{Introduction}
In 1993 R. Camassa and D. Holm \cite{CH} proposed a new equation
\begin{equation}\label{ch}
{\tt u}_t + 3{\tt u} {\tt u}_{x} = ({\tt u}_{xxt}+2{\tt u}_x{\tt
u}_{xx}+{\tt u}{\tt u}_{xxx})-2\nu {\tt u}_x,
\end{equation}
with $\nu$ a constant parameter, deriving it as the governing
equation for waves in shallow water when surface tension is
present. This involves an asymptotic expansion in small amplitude
of the incompressible Euler equation for unidirectional motion
under the influence of gravity that extends one order beyond the
Korteweg-de Vries (KdV) equation.  (\ref{ch})  is also an element
in a class of equations introduced by A.Fokas and B.Fuchssteiner
\cite{FF} through the method of recursion operators in 1981.

Equation (\ref{ch}) is strongly nonlinear, admits a bi-Hamiltonian
structure \cite{FF}, a Lax pair \cite{CH} and it is formally
integrable through the inverse scattering method \cite{Con}. The
bi-Hamiltonian structure of the CH equation can be described as
follows
\[
m_t=P_1\displaystyle\frac{\delta H_2}{\delta m},\;\;\;\;
m={\tt u}-{\tt u}_{xx},\quad P_1=-\partial_x+\partial_x^3,
\]
with
\begin{equation}
\label{H2} H_2=\frac12\int ({\tt u}^3+{\tt u}{\tt u}_x^2+2\nu {\tt u}^2)dx,
\end{equation}
or
\[
m_t=P_2\displaystyle\frac{\delta H_1}{\delta m}, \quad
P_2=-\partial_xm-m\partial_x-2\nu\partial_x,
\]
and
\begin{equation}
\label{H1} H_1=\frac12\int ({\tt u}^2+{\tt u}_x^2)dx.
\end{equation}
The bi-Hamiltonian structure implies that the  CH equation has an
infinite number of conserved quantities. The functionals $H_k$,
$k\in \mathbb{Z},$ defined by
\begin{equation}
\label{Hn} P_2\displaystyle\frac{\delta H_k}{\delta
m}=P_1\displaystyle\frac{\delta H_{k+1}}{\delta m},\quad k\in
\mathbb{N},
\end{equation}
are conserved quantities in involution with respect to the Poisson
bracket determined either by $P_1$ or $P_2$. The Hamiltonian
$H_0=\int mdx $ is the  Casimir of the first Poisson tensor $P_1$.
For $k>2$ the Hamiltonian densities of  $H_k$ are not local
functions of ${\tt u}$ and its spatial derivatives.

In the case $\nu=0$, R. Camassa and D. Holm \cite{CH} proved the
existence of solutions that are continuous but only piece-wise
analytic (peakons). The CH equation possesses soliton solutions,
periodic finite-gap solutions \cite{Con1,CM,AlFe},  real
finite-gap solutions \cite{FrH} and, for $\nu=0$, multi-peakons
\cite{BSS}. In particular, the algebro--geometric solutions of
(CH) are described as Hamiltonian flows on nonlinear subvarieties
(strata) of generalized Jacobians. This implies that the
associated finite dimensional integrable systems may be described
in the framework of integrable systems with deficiency
\cite{Van,AF} whose algebraic--geometrical structure has much in
common with the celebrated algebraically completely integrable
systems introduced and thoroughly studied by M. Adler and P. van
Moerbeke \cite{AvM}.

In this work we derive  the Whitham  modulation equations for the
CH flows. Whitham modulation equations for a nonlinear evolution
system describe slow modulations of parameters over a family of
periodic travelling wave solutions (or families of multi--phase
solutions which are so far known to exist only for integrable
systems). Contrary to the Korteweg de Vries case \cite{LL}, it is
an open problem to show that the Cauchy problem for CH with slowly
varying initial data is described by the Whitham equations. Both
for KdV and CH equations, this approximation is physically
meaningful when the ratio between the water depth and the
wavelength is very small \cite{DGH}.

The Whitham equations  are a system of hydrodynamic type equations
\cite{DN} and in the Riemann invariant coordinates take the  form
\[
u^i_t+v^i(\vu)u^i_x=0, \quad i=1,\dots,N, \quad \vu =
(u^1,\dots,u^N),
\]
where we denote fast and slow variables with the same letters $x$ and $t$ and upper
indeces denote controvariant  vectors.
The original evolution system is usually Lagrangian or Hamiltonian
and this property is usually inherited by the equations of slow
modulations. To average the original equations in the Lagrangian
form, Whitham \cite{W} introduced the pseudo-phase method and then
he constructed the corresponding Hamiltonian structure. For local
Hamiltonian structures, B.A. Dubrovin and S.P. Novikov \cite{DN}
introduced a procedure for averaging local Poisson brackets and
obtained the corresponding modulation equations. A third method to
derive the Whitham modulation equations is the nonlinear analog of
the WKB method \cite{DM}. It can be proven that the three methods
lead to the same equations for the case in which the original
equation has a local Hamiltonian structure and local Hamiltonian
densities (see \cite{DN} and references therein).

The Camassa-Holm equation does not possess  a local Hamiltonian
structure: indeed in the variable ${\tt u}$ the Hamiltonian operator is
strongly nonlocal and  the Hamiltonian densities of $H_k$ are
non-local for $k>2$. In the variable $m$ the Hamiltonian densities
of $H_k$ are  nonlocal for $k>0$. Therefore  the CH equation does
not fit into the method of averaging local Hamiltonian structure
\cite{DN} nor even in the Maltsev-Novikov method of averaging
weakly nonlocal Hamiltonian structures \cite{MN} (an Hamiltonian
structure is weakly nonlocal if it is polynomial in $\partial_x$
and its higher derivatives and linear in $\partial_x^{-1}$) or in
the Maltsev method \cite{Ma} of averaging weakly non-local
symplectic form (inverse of the Hamiltonian operator). The latter
method applies to Camassa-Holm only when averaging one--phase
solutions \cite{Maa}.

The CH  equation can  be written as a local Lagrangian system and
we use the Whitham method (modulation equations in Lagrangian
form) to derive the modulation equations for the one-phase
periodic solution.  The CH modulation equations for the Riemann
invariants $u^1<u^2<u^3$, take the form
\[
\partial_t u^i + C^i({\vu}) \partial_x u^i=0, \quad\quad i=1,\dots,3,
\]
where
\begin{equation*}
%\label{Ci1}
\begin{split}
&C^1(u^1,u^2,u^3)=u^1+u^2+u^3+2\nu+2\displaystyle\frac{(u^1+\nu)(u^1-u^2)
\Lambda(K(s),\rho,s)}
{(u^2+\nu)[K(s)-E(s)]}
\\
&C^2(u^1,u^2,u^3)=u^1+u^2+u^3+2\nu+\displaystyle\frac{2(u^2-u^1)\Lambda(K(s),
\rho,s)}{K(s)-\displaystyle\frac{(u^2+\nu)(u^3-u^1)}{(u^1+\nu)(u^3-u^2)}E(s)}
\\
&C^3(u^1,u^2,u^3)=u^1+u^2+u^3+2\nu+2\displaystyle\frac{(u^1+\nu)(u^3-u^2)
\Lambda(K(s),\rho,s)}{(u^2+\nu)E(s)}.
\end{split}
\end{equation*}
In the above formulas $K(s)$ and $E(s)$ are the complete elliptic
integrals of the first and
 second  kind with modulus
$s^2=\displaystyle\frac{(u^2-u^1)(u^3+\nu)}{(u^3-u^1)(u^2+\nu)}$
and $\Lambda(K(s),\rho,s)$ is the complete elliptic integral of
the third kind defined by
\[
\Lambda(K(s),\rho,s)=\int_0^{K(s)}\displaystyle\frac{d v}{
1-\rho^2 sn^2 v},\quad
\rho^2=\displaystyle\frac{u^2-u^1}{u^2+\nu},
\]
with $sn$ the Jacobi elliptic function. The equations are
hyperbolic for $-\nu<u^1<u^2<u^3$ where $\nu$ is the parameter
entering in the CH equation (\ref{ch}).

Then following Hayes \cite{H} and Whitham
\cite{W} the equations can be written in Hamiltonian form with a
local Poisson bracket of Dubrovin-Novikov type
\[
u^i_t=-C^i(\vu)u^i_x=A^{ij}\displaystyle\frac{\partial h}{\partial
u^j}
\]
where
\begin{equation}
\label{H0}
A^{ij}=g^{ii}\delta^{ij}\displaystyle\frac{d}{dx}-g^{ii}\Gamma^j_{ik}u^k_x
\end{equation}
is the Hamiltonian operator and $h$ the Hamiltonian density. As
pointed out by Dubrovin and Novikov, $A^{ij}$ defines a
Hamiltonian operator if and only if $g^{ii}=g^{ii}(\vu)$ is a flat
non degenerate metric and $\Gamma^j_{ik}$ are the Christoffel
symbols of the corresponding Levi-Civita connection. We also find
a second local compatible Hamiltonian structure which is obtained
from the flat metric $g^{ii}(\vu)(u^i+\nu)$ where $\nu $ is the
constant in the CH equation. Therefore the nonlocal bi-Hamiltonian
structure of the original CH equation averages to a local
bi-Hamiltonian structure of Dubrovin Novikov type.

\smallskip

A reciprocal transformation is a closed form which changes the
independent variables of the equation and maps conservation laws
into conservations laws, but it does not preserve the Poisson
structure as shown by E.V. Ferapontov and M.V. Pavlov
\cite{FP},\cite{ferapontov}. The Camassa Holm equation can be
transformed by a reciprocal transformation into the first negative
flow of the  KdV hierarchy \cite{Fu} (also known as AKNS equation
\cite{AKNS}). An elegant treatment  of the relations among
positive and negative flows of the CH and KdV hierarchies can be
found in \cite{MK}.

Let $g^{KdV}_{ii}$ and $g^{KdV}_{ii}/\beta^i$ be the flat
compatible metrics associated to the bi-Hamiltonian structure of
the KdV modulation equations \cite{T}\cite{Du1} with respect to
the usual Riemann  invariants $\beta^1,\beta^2,\beta^3$ as defined
in \cite{W}. Then the reciprocal transformation is generated by
the Casimir  $\mathcal{H}_0$ of the Hamiltonian  operator
associated to the metric $g^{KdV}_{ii}/\beta^i$. According to the
results in \cite{FP},\cite{ferapontov}  the reciprocal
transformation maps the two KdV flat metrics to the CH metrics
\[
\displaystyle\frac{g^{KdV}_{ii}}{{\cal H}^2_0},\quad
\displaystyle\frac{g^{KdV}_{ii}}{{\cal H}^2_0\beta^i}
\]
which are not flat. The relation between the CH Riemann invariants
and the KdV Riemann invariants is $\beta^i=1/(u^i+\nu)$. The
corresponding CH modulation equations are Hamiltonian with respect
to two non-local operators of Mokhov-Ferapontov and Ferapontov
type \cite{FM},\cite{ferapontov} which are of the form (\ref{H0})
plus a nonlocal tail. However from the Lagrangian averaging, we
independently prove the existence of one local Hamiltonian
structure. We show that the two metrics
\[
\displaystyle\frac{g^{KdV}_{ii}}{{\cal H}^2_0(\beta^i)^2},\quad
\displaystyle\frac{g^{KdV}_{ii}}{{\cal H}^2_0(\beta^i)^3}
\]
are flat and define a flat pencil, that is, the  CH modulation
equations are bi-Hamiltonian with respect to two local Hamiltonian
operators of the form (\ref{H0}). We remark that the two flat KdV
metrics $g^{KdV}_{ii}$ and $g^{KdV}_{ii}/\beta^i$ are related to a
semisimple Frobenius manifold \cite{Du2}. More in general, B.
Dubrovin \cite{Du2} proves that, under certain assumptions, a flat
pencil of contravariant metrics on a manifold induces a Frobenius
structure on it. One of the assumptions is the requirement that
one of the two flat metrics is of Egorov type (namely its rotation
coefficients are symmetric). Since none of the two CH flat metrics
have the Egorov property, there is no Frobenius structure
associated to this system. Therefore, from the geometric point of
view, the KdV modulation equations and the CH modulation equation
belong to two different classes.

All the results presented here for the one-phase CH modulation
equations may be generalized to the multi--phase case in a
straightforward way. However, in the present paper we have decided
to concentrate only on the one-phase case to better clarify
similarities and differences with the KdV case, and we will
present the discussion of the multi--phase case in a future
publication.

\section{Whitham modulation equations}
In this section we use Lagrangian formalism to average the
Camassa-Holm equation in the genus one case and refer to
\cite{DN},\cite{MaPa} for a general exposition of the method we use.
Introducing the potential
\[
\phi \; : \quad \phi_x = {\tt u},
\]
equation (\ref{ch}) takes the form
\[
\phi_{xt} -\phi_{xxxt} + 3\phi_x\phi_{xx} -2\phi_{xx}\phi_{xxx} -
\phi_x\phi_{xxxx} +2\nu\phi_{xx}
=0,
\]
and a Lagrangian is
\begin{equation}
\label{lagch}
\mathcal{L}= -\frac{1}{2}
\phi_x\phi_t +\frac{1}{2}\phi_{xxx}\phi_t -\frac{1}{2}\phi_x^3
-\nu\phi_x^2 +\frac{1}{4}\phi_x^2\phi_{xxx}.
\end{equation}
We consider $2\pi$--periodic solutions of the form
\[
{\tt u}=\eta(\theta),\quad \theta=kx-\omega t.
\]
Following Whitham \cite{W}, we introduce the pseudo-phase
\[
\phi =\psi +\Phi(\theta),\;\;  \psi=\beta x -\gamma t, \quad\theta=kx-\omega t,
\]
where $\Phi(\theta)$ is  a $2\pi$ periodic function of $\theta$
with zero average. The averaged Lagrangian over the
one-dimensional real torus is
\begin{align*}
\bar{\mathcal{L}}&\;=\;\displaystyle\oint d\theta
\left[-\frac12(\beta+k\Phi_{\theta})(\gamma-\omega\Phi_{\theta})+
\frac{1}{2}k^3\Phi_{\theta\theta\theta}
(\gamma-\omega\Phi_{\theta}) -\right.\\
&\;\;\left.-\frac{1}{2}(\beta+k\Phi_{\theta})^3-\nu(\beta+k
\Phi_{\theta})^2+\displaystyle\frac14(\beta+k\Phi_{\theta})^2k^3
\Phi_{\theta\theta\theta}\right]
\end{align*}
Now we suppose that the constants $\beta$, $k$, $\gamma$ and
$\omega$ are slowly varying functions of time, that is
$\beta=\beta(X,T)$, $\gamma=\gamma(X,T)$, $k=k(X,T)$ and
$\omega=\omega(X,T)$ with $X$ and $T$ ``slow variables''. Then
\[
\bar{\mathcal{L}}=\bar{\mathcal{L}}(k,\omega,\beta,\gamma;X,T).
\]
The equations of slow modulation of the parameters
$k,\omega,\beta,\gamma$ are the extremals of the functional \cite{W}
\[
\int\bar{\mathcal{L}}(k,\omega,\beta,\gamma)dX\,dT,
\]
and take the form
\begin{equation}
\label{mod3}
\left\{\begin{array}{c}
\partial_{X}\bar{\mathcal{L}}_{k}+\partial_{T}\bar{\mathcal{L}}_{\omega}=0,
\quad k_T+\omega_X=0,\\
\partial_{X}\bar{\mathcal{L}}_{\gamma}+\partial_{T}\bar{\mathcal{L}}_{\beta}=0,
\quad \beta_T+\gamma_X=0.
\end{array}\right.
\end{equation}
As in the KdV case, substituting the forth equation into the third
in (\ref{mod3}), the $X$-derivative disappears and we get a
constraint. Thus the number of equations reduces from four to
three.

We get to the same conclusion transforming the modulation
equations in Hamiltonian form. Following Hayes \cite{H} and
Whitham \cite{W}, let us introduce the Hamiltonian density
\[
\mathcal{H}=\mathcal{H}(k,\omega,\beta,\gamma)=
\omega\bar{\mathcal{L}}_{\omega}+\gamma\bar{\mathcal{L}}_{\gamma}-\bar{\mathcal{L}}.
\]
Then the modulation equations are Hamiltonian with respect to the
canonical Poisson bracket
\[
\{k(X),\bar{\mathcal{L}}_{\omega}(Y)\}=\delta'(X-Y),
\quad \{\beta(X),\bar{\mathcal{L}}_{\gamma}(Y)\}=\delta'(X-Y).
\]
Since the constraint
\[
\bar{\mathcal{L}}_{\gamma}=\displaystyle\frac{1}{2}\beta,
\]
the number of fields reduces from four to three. This is connected with the
Dirac reduction. Therefore the Whitham equations can be written in
Hamiltonian form with a local Dubrovin-Novikov Poisson bracket
\cite{Maa}
\[
\{k(X),\bar{\mathcal{L}}_{\omega}(Y)\}=\delta'(X-Y),
\quad \{\beta(X),\beta(Y)\}=2\delta'(X-Y)
\]
and with Hamiltonian
$\mathcal{H}=\mathcal{H}(k,J,\beta)$, $J=\bar{\mathcal{L}}_{\omega}$.
In these variables the equations of motion are Hamiltonian
\begin{equation}
\label{modH}
k_T=\partial_X \mathcal{H}_J,\quad J_T=\partial_X \mathcal{H}_k,\quad \beta_T=
\partial_X \mathcal{H}_{\beta}.
\end{equation}

\section{Modulation equations for CH in Riemann invariant form}
The modulation equations (\ref{modH}) can also be written in
Riemann invariant form. For the purpose we introduce the spectral
curve associated to the periodic travelling wave solution
${\tt u}(x,t)=\eta(kx-\omega t)$. When we plug $\eta(\theta)$, $\theta=kx-\omega
t$, into the CH equation (\ref{ch}), we get, after integration,
\begin{equation}
\label{periodic2} k^2(c- \eta )\eta^2_{\theta}+ (2\nu-c)\eta^2+\eta^3+2B\eta-2A=0,
\end{equation}
where $A$ and $B$ are constants of integration and $c=\omega/k$.
The CH one-phase solution ${\tt u}(x,t)=\eta(kx-\omega t)$
is obtained by inverting the third kind differential
\[
\int_{u_0}^u\displaystyle\frac{(\eta-c) d\eta}{
\sqrt{(\eta-c)(\eta^3+ (2\nu-c)\eta^2 +2B\eta-2A)}}=kx-\omega t.
\]
The inversion of the above integral is discussed in \cite{AlFe}
where the solution ${\tt u}(x,t)$ is expressed in terms of convenient
generalized theta-functions in two variables, which are
constrained to the generalized theta-divisor. Integration of
(\ref{periodic2}) over $\theta$ yields the amplitude dependent
dispersion relation for the nonlinear dispersive wave
\begin{equation}
\label{k} k \oint\displaystyle\frac{(\eta-c) d\eta}{
\sqrt{(\eta-c)(\eta^3+ (2\nu-c)\eta^2 +2B\eta-2A)}}= 2\pi,
\end{equation}
where the  integration is taken on a closed path between $e^2$ and
$e^1$ where $c>e^1>e^2>e^3$  are the roots of the polynomial
\[
\eta^3 -(c-2\nu)\eta^2 +2B \eta -2A,
\]
with the constraint
\[
2\nu=c-e^1-e^2-e^3.
\]
Here and below, we denote vectors with upper indices.

>From now on, we will use small letters $x,t$ for the `slow
variables' $X, T$ introduced in the previous section, since we
deal only with modulation equations and no ambiguity may occur.

\begin{figure}[htb]
\centering
\mbox{\epsfig{figure=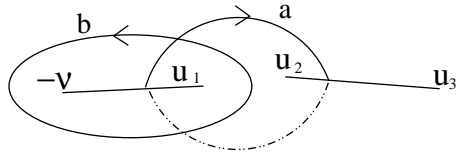,width=0.4\textwidth}}
     \caption{The homology basis. \label{fig1}}
\end{figure}

\begin{theorem}
\label{theoremRI}
The one-phase CH modulation  equations (\ref{modH}) take the
Riemann invariant form
\begin{equation}\label{whithamu}
\partial_t u^i + C^i({\vu}) \partial_x u^i=0, \quad\quad i=1,\dots,3,
\end{equation}
where the Riemann invariants $\vu=(u^1,u^2,u^3)$, $u^1<u^2<u^3$,  are
\[
u^1=\frac{1}{2}(e^2+e^3), \quad u^2=\frac{1}{2}(e^1+e^3), \quad
u^3=\frac{1}{2}(e^1+e^2), \quad
\]
and the speeds $C^i(\vu)$ take the form
\begin{equation}
\label{velu}
C^i(\vu)=\displaystyle\frac{\partial_{u^i}\omega(\vu)}{\partial_{u^i}k(\vu)}.
\end{equation}
The wave number and frequency are given by the Abelian integrals
\begin{equation}
\begin{split}
\label{komega}
k&=2\pi\left(\oint_a\displaystyle\frac{(\lambda+\nu)d\lambda}{\sqrt{R(\lb)}}
\right)^{-1},
\;\; R(\lambda)=(\lambda+\nu)(\lb-u^1)(\lb-u^2)(\lb-u^3),\\
\omega&=(2\nu+u^1+u^2+u^3)k,
\end{split}
\end{equation}
and the  integration is taken on cycle  $a$  passing between $u^2$
and $u^1$ (see figure~\ref{fig1}).
\end{theorem}

The existence of the Riemann invariants and equations (\ref{velu})
is proven directly starting from (\ref{modH}) and  using
variational identities of Abelian integrals. To write the velocities
in an explicit form, we introduce the integrals
\begin{equation}
\label{Ik} I_k = \oint_{a}  \frac{\lambda^k}{\sqrt{R(\lambda) }}
d\lambda, \quad k\ge 0,
\end{equation}
where
$R(\lambda)$ is defined in (\ref{komega}).

Let $\sigma_1(\lambda)$ be the normalized third kind differential
with first order pole at $(\infty,\pm\infty)$ with residue $\pm
1$, respectively, and let $\sigma_2$ be the normalized second kind
differential with second order pole at infinity, namely
\begin{align}
\label{P1} &\sigma_1(\lambda)=\displaystyle\frac{P_1
(\lambda)d\lambda}{\sqrt{R(\lambda)}},\quad P_1(\lambda)
= \lambda +\gamma_1, \\
\label{P2} &\sigma_2(\lambda) = \displaystyle\frac{P_2
(\lambda)d\lambda}{\sqrt{R(\lambda)}},\quad P_2 (\lambda) =
\lambda^2 -\displaystyle\frac{1}{2}(u^1+u^2+u^3-\nu) \lambda +
\gamma_2
\end{align}
where  the constants $\gamma_1=-\displaystyle\frac{I_1}{I_0}$ and
$\gamma_2= -\displaystyle\frac{I_2}{I_0}+
\displaystyle\frac{1}{2}(u^1+u^2+u^3-\nu)\displaystyle\frac{I_1}{I_0}$
are uniquely determined by
\begin{equation}
\label{normsigma}
\displaystyle
\oint_{a}\sigma_i(\lambda)=0,\quad i=1,2.
\end{equation}
These constants are explicitly given by
\begin{equation}
\label{gamma1} \gamma_2 =\displaystyle\frac{1}{2}[u^1u^2-\nu
u^3+(u^3-u^1)(u^2+\nu)\displaystyle\frac{E(s)}{K(s)}], \quad
\gamma_1 =
\nu-(u^1+\nu)\displaystyle\frac{\Lambda(K(s),\rho,s)}{K(s)},
\end{equation}
where
\[
K(s)=\int_0^{\pi/2}\displaystyle\frac{d\psi}{
\sqrt{1-s^2\sin^2\psi}},\quad E(s)=\int_0^{\pi/2}d\psi
\sqrt{1-s^2\sin^2\psi}
\]
 are the complete elliptic integrals of the first and second kind
respectively with modulus
\[
s^2=\displaystyle\frac{(u^2-u^1)(u^3+\nu)}{(u^3-u^1)(u^2+\nu)}
\]
and
\[
\Lambda(K(s),\rho,s)=\int_0^{K(s)}\displaystyle\frac{d v}{
1-\rho^2 sn^2 v},\quad
\rho^2=\displaystyle\frac{u^2-u^1}{u^2+\nu},
\]
is the complete elliptic integral of the third kind with $sn$ the
Jacobi elliptic function.
\begin{theorem}
The speeds $C^i(\vu)$, $i=1,2,3$ defined in (\ref{velu}) take the
form
\begin{equation}
\label{Ci1}
\begin{split}
&C^1(u^1,u^2,u^3)=u^1+u^2+u^3+2\nu+2\displaystyle\frac{(u^1+\nu)(u^1-u^2)
\Lambda(K(s),\rho,s)}{(u^2+\nu)[K(s)-E(s)]}
\\
&C^2(u^1,u^2,u^3)=u^1+u^2+u^3+2\nu+\displaystyle\frac{2(u^2-u^1)
\Lambda(K(s),\rho,s)}{K(s)-
\displaystyle\frac{(u^2+\nu)(u^3-u^1)}{(u^1+\nu)(u^3-u^2)}E(s)}
\\
&C^3(u^1,u^2,u^3)=u^1+u^2+u^3+2\nu+2\displaystyle\frac{(u^1+\nu)(u^3-u^2)
\Lambda(K(s),\rho,s)}{(u^2+\nu)E(s)}.
\end{split}
\end{equation}
where $K(s)$, $E(s)$ and $\Lambda(K(s),\rho,s)$ are the complete
elliptic integrals of first, second and third kind with modulus
$s^2=\displaystyle\frac{(u^2-u^1)(u^3+\nu)}{(u^3-u^1)(u^2+\nu)}$.

\noindent The equations  $\partial_t u^i + C^i(\vu) \partial_x
u^i=0$ are hyperbolic and the velocities satisfy
\[
C^1(\vu)<C^3(\vu),\;\;C^2(\vu)<C^3(\vu),\quad -\nu<u^1<u^2<u^3.
\]
In the limit when  two Riemann invariants coalesce, the modulation
equation reduce to the dispersionless CH equation
\[
\partial_t{\tt u} + (3{\tt u}+2\nu)\partial_x {\tt u}=0.
\]
\end{theorem}
To prove the theorem we introduce the normalized holomorphic differential
$\phi(\lambda)$
\[
\phi(\lambda)=\displaystyle\frac{d\lambda}{I_0\sqrt{R(\lambda)}},\quad
\oint_{a}\phi(\lambda)=1.
\]
Next we observe that the wave number $k$ defined
in (\ref{komega}) takes the form
\begin{equation}
\label{kdiff} k=-2\pi
\displaystyle\frac{\phi(-\nu)}{\sigma_1(-\nu)},
\end{equation}
where
\[
\sigma_1(-\nu):=\displaystyle\frac{2P_1(-\nu)}{\sqrt{(-\nu-u^1)(-\nu-u^2)(-\nu-u^3)}}=
\left.\displaystyle\frac{\sigma_1(\lambda)}{dt}\right|_{\lambda=-\nu},\quad
t^2=\lambda+\nu,\] and
\[
\phi(-\nu):=\displaystyle\frac{2}{I_0\sqrt{(-\nu-u^1)(-\nu-u^2)(-\nu-u^3)}}=\left.
\displaystyle\frac{\phi(\lambda)}{dt}\right|_{\lambda=-\nu},\;\;
t^2=\lambda+\nu,
\]
with   $\sigma_1(\lambda)$ defined in
(\ref{P1}). The following variational formulas
hold \cite{Ko}
\begin{equation}
\label{diff1} \displaystyle\frac{\partial}{\partial
u^i}\phi(-\nu)=
\displaystyle\frac{1}{2}\phi(u^i)\Omega_{\nu}(u^i),\quad\quad\quad
\displaystyle\frac{\partial}{\partial
u^i}\sigma_1(-\nu)=\displaystyle\frac{1}{2}\sigma_1(u^i)\Omega_{\nu}(u^i),
\end{equation}
where
\begin{equation}
\label{hatomega} \phi(u^i):=
\left.\displaystyle\frac{\phi(\lambda)}{dt}\right|_{\lambda=u^i},\,\,
\Omega_{\nu}(u^i):=\left.\displaystyle\frac{\Omega_{\nu}(\lambda)}{dt}
\right|_{\lambda=u^i},\quad t^2=\lambda-u^i.
\end{equation}
and $\Omega_{\nu}(\lambda)$   is a second kind Abelian
differential with second order pole at $\lambda=-\nu$ with
asymptotic behavior for $\lambda\rightarrow-\nu$
\[
\Omega_{\nu}(\lambda) \rightarrow \displaystyle\frac{dt}{t^2},
\quad t^2=\lambda+\nu,
\]
and normalized by the condition
\begin{equation}
\label{normOm}
\oint_{a}\Omega_{\nu}(\lambda)=0.
\end{equation}
The differential $\Omega_{\nu}(\lambda)$ is explicitly given by the expression
\begin{equation}
\begin{split}
\label{omeganu}
\Omega_{\nu}(\lambda)&=\displaystyle\frac{P_{\nu}(\lambda)d\lambda}{\sqrt{R(\lambda)}
\sqrt{(-\nu-u^1)(-\nu-u^2)(-\nu-u^3)}},\\
P_{\nu}(\lambda)&=
\displaystyle\frac{-(\nu+u^1)(\nu+u^2)(\nu+u^3)}{2(\lambda+\nu)}+P_2(-\nu),
\end{split}
\end{equation}
where $P_2(\lambda)$ has been defined in (\ref{P2}).
Inserting (\ref{kdiff}) and (\ref{diff1}) into (\ref{velu}), we
finally obtain the expression
\begin{equation}
\label{Ci} C^i(\vu)=u^1+u^2+u^3+2\nu
-\frac{P_1(-\nu)}{P_{\nu}(u^i)}\prod_{j\neq i,
j=1}^3(u^i-u^j),\;i=1,2,3,
\end{equation}
where the rational function $P_1(\lambda)$ and $P_{\nu}(\lambda)$
are as in (\ref{P1}) and (\ref{omeganu}) respectively.
Finally inserting (\ref{gamma1}) into (\ref{Ci})  we arrive at the formula
(\ref{Ci1}).

\noindent Next we prove the ordering of the velocities using the
formula (\ref{Ci}).  Since $-\nu<u^1<u^2<u^3$, and $\displaystyle
\lim_{ \lambda\rightarrow- \nu^+}P_{\nu}(\lambda)=-\infty$,  by
monotonicity, there is only one point $\lambda^*>-\nu$ for which
\[
\displaystyle\frac{(\nu+u^1)(\nu+u^2)(\nu+u^3)}{2(\lambda^*+\nu)}=P_2(-\nu);
\]
moreover, because of (\ref{normOm}), the point $\lambda^*$
satisfies the inequality $u^1<\lambda^*<u^2$ . Therefore
\[
P_{\nu}(u^1)<0,\quad P_{\nu}(u^3)> P_{\nu}(u^2)>0.
\]
In the same way, using (\ref{normsigma}) we conclude that
$P_1(-\nu)<0$. Using the above inequalities, it is straightforward
to verify that
\[
C^1(\vu)<C^3(\vu),\;\;\;,C^2(\vu)<C^3(\vu),\quad -\nu<u^1<u^2<u^3.
\]
In general there is not a strict ordering between 
$C^1(\vu)$ and $C^2(\vu)$. For example  for step-like initial data
the characteristics $C^1(\vu)$ and $C^2(\vu)$ do cross for $u^1<u^2<u^3$.

We end the proof of the theorem studying the behavior of the
speeds $C^i(\vu)$,  $i=1,2,3$, when two of the Riemann invariants
coalesce. In the limiting case
\[
u^2=v-\epsilon,\quad u^3=v+\epsilon,\quad\epsilon\rightarrow 0,\]
  we have
\[
\lim_{\epsilon\rightarrow 0}P_1(\lambda)=\lambda-v,\quad
\lim_{\epsilon\rightarrow 0} P_{\nu}(
\lambda)=\frac{1}{2(\lambda+\nu)}(\lambda-v)(u^1+\nu)(v+\nu),
\]
so that
\[
\quad\lim_{\epsilon\rightarrow
0}\frac{P_1(-\nu)}{P_{\nu}(u^1)}(u^1-u^2)(u^1-u^3)=2( v-u^1)
\]
and
\[
\quad\lim_{\epsilon\rightarrow 0}\frac{P_1(-\nu)}{P_{\nu}(u^2)}(u^2-u^1)(u^2-u^3)
=\lim_{\epsilon\rightarrow 0}\frac{P_1(-\nu)}{P_{\nu}(u^3)}(u^3-u^2)(u^3-u^1).
\]
Combining the above relations, we have
\[
C^2(u^1,v,v)=C^3(u^1,v,v),\quad
C^1(u^1,v,v)=3u^1+2\nu,
\]
which gives the dispersionless CH equation $\partial_t
u^1+(3u^1+2\nu)\partial_x u^1=0$. In the same way, it can be
proved that, in the limit $u^2=u^1$, the speeds
$C^1(u^1,u^1,u^3)=C^2(u^1,u^1,u^3)$ and $C^3(u^1,u^1,u^3)=
3u^3+2\nu$.

\subsection{Hamiltonian structure and integration}
In this section, we investigate the bi-Hamiltonian structure of
the one-phase Whitham equations
\[
\partial_t u^i + C^i(\vu) \partial_x u^i =0, \quad i=1,\dots,
3,
\]
with $C^i(\vu)$ as in (\ref{velu}) or (\ref{Ci}). In section 2
we have proven that the above equations are Hamiltonian with
respect to a canonical Poisson bracket. In the following  we show
that the CH modulation equations are bi-Hamiltonian with respect
to local Poisson brackets of Dubrovin-Novikov type.
\begin{proposition}
\label{metricu}
The speeds $C^i(\vu)$, $i=1,2,3,$ satisfy the following relations
\begin{equation}
\label{tsarev} \displaystyle\frac{\partial_{u^j} C^i }{C^j-C^i}
=\partial_{u^j}\log \sqrt{g_{ii}},
\end{equation}
where $g_{ii}=g_{ii}(\vu)$ is the covariant metric defined by the relation
\begin{equation}\label{f1CH}
g_{ii}(\vu)=-4(u^i+\nu)\displaystyle\frac{\res[\lambda=u^i]
\left\{\displaystyle\frac{(\Omega_{\nu}(
\lambda))^2 }{d\lambda}
\right\}}{\res[\lambda=-\nu]\left\{\displaystyle\frac{(\sigma_1(\lambda))^2}{
d\lambda}\right\}},
\end{equation}
with $\Omega_{\nu}(\lambda)$ the second kind  differential defined
in (\ref{omeganu}) and $\sigma_1(\lambda)$ the third kind
differential defined in (\ref{P1}). The metric $g_{ii}(\vu)$ is
flat.   The metric $g_{ii}(\vu)$ is defined up to multiplication by an arbitrary
function $f_i(u^i)$. The metrics
\begin{equation}
\label{f2CH} \displaystyle \frac{g_{ii}(\vu)}{2(u^i+\nu)},\quad
\displaystyle \frac{g_{ii}(\vu)}{4(u^i+\nu)^2},\quad \displaystyle
\frac{g_{ii}(\vu)}{8(u^i+\nu)^3},
\end{equation}
are respectively flat, of constant curvature $R^{ij}_{ij}=-1$  and
conformally flat with curvature tensor
$R^{ij}_{ij}=-C^i(\vu)-C^j(\vu)-2\nu$. The pencil of metrics
\begin{equation}
\label{pencil} g_{ii}(\vu)+\lambda
\displaystyle\frac{g_{ii}(\vu)}{(u^i+\nu)}
\end{equation}
is flat for any real $\lambda$. None of the metrics defined in
(\ref{f1CH}) and in (\ref{f2CH}) is of Egorov type.
\end{proposition}
We denote the diagonal metric in covariant form by $g_{ii}$  and
its inverse by $g^{ii}$.
To derive (\ref{tsarev}) it is sufficient to use the variational formulas
(\ref{diff1}) and the additional one
\begin{equation*}
\label{diff2} \displaystyle\frac{\partial}{\partial
u^i}\Omega_{\nu}(u^j)=
\displaystyle\frac{1}{2}\Omega_{\nu}(u^i)\Omega_{u^i}(u^j),\quad\quad\quad
\end{equation*}
where $\Omega_{u^i}(\lambda)$ is a normalized
second kind differential with second order pole at $\lambda=u^i$.
The explicit form of $\Omega_{u^i}(\lambda)$ can be obtained from
(\ref{omeganu}) by replacing
$-\nu$ by $u^i$. The quantities $\Omega_{\nu}(u^i)$ and
$\Omega_{u^i}(u^j)$ are defined as in (\ref{hatomega}).

To prove the second part of the proposition we evaluate the
non-zero elements of the curvature tensor $R^{ij}_{il}$
\begin{align}
\label{rot1}
R^{ij}_{il} & =- \frac{1}{\sqrt{g_{ii} g_{jj}}} \left\{ \partial_l
r_{ji}-r_{jl} r_{li}\right\}\,,i\neq j\neq l,\\
\label{rot2}
R^{ij}_{ij} & = - \frac{1}{\sqrt{g_{ii} g_{jj}}} \left\{
\partial_i r_{ij} + \partial_j r_{ji} +
\sum_{p\neq i,j} r_{pj} r_{pi} \right\}\,.
\end{align}
Here  $r_{ij}$ are the rotation coefficients defined by
\begin{equation}
\label{rotation}
r_{ij}=\displaystyle\frac{\partial_{u^i}\sqrt{g_{jj}}}{\sqrt{g_{ii}}},\quad
i\neq j.
\end{equation}
A metric is of Egorov type if $r_{ij}= r_{ji}$, $\forall i \neq
j$. By direct calculation we obtain \[
r_{ij}=\displaystyle\frac{1}{2}\left(\displaystyle\frac{u^j+\nu}{u^i+\nu}
\right)^{k/2}\left(\Omega_{u^i}(u^j)-
\displaystyle\frac{\Omega_{\nu}(u^j)\sigma_1(u^i)}{\sigma_1(-\nu)}\right),\quad
k=1,0,-1,-2 .
\]
>From the above formula  it is clear that $r_{ij}\neq r_{ji}$, for
all the four metrics and  therefore none of them is an Egorov
metric. To evaluate (\ref{rot1}-\ref{rot2}) the following
additional variational formulas are needed
\begin{equation}
\label{diff0}
\begin{split}
\displaystyle\frac{\partial}{\partial
u^i}\gamma_1&=-\displaystyle\frac{1}{2}+\displaystyle\frac{1}{4}\sigma_1(u^i)
\sigma_2(u^i),\;\;i=1,2,3,\\
\displaystyle\frac{\partial}{\partial
u^i}\gamma_2&=\displaystyle\frac{1}{4}(u^1+u^2+u^3-\nu)-\displaystyle\frac{1}{2}u^i+
\displaystyle\frac{1}{4}(\sigma_2(u^i))^2,\;\;i=1,2,3,
\end{split}
\end{equation}
where $\sigma_1$ and $\sigma_2$ have been defined in (\ref{P1})
and (\ref{P2}) respectively and
\[
\sigma_k(u^i):=\left.\displaystyle\frac{\sigma_k(\lambda)}{dt}\right|_{\lambda=u^i},\quad
t^2=\lambda-u^i,\quad i=1,2,3,\;k=1,2.
\]
Using the variational formulas (\ref{diff0}) we obtain that
$R^{ij}_{il}=0$ for all the four metrics and
\begin{equation}
\nonumber
\begin{split}
g_{ii}(\vu)&\rightarrow R_{ij}^{ij}=0,\\
\displaystyle\frac{g_{ii}(\vu)}{2(u^i+\nu)}&\rightarrow R_{ij}^{ij}=0,\\
\displaystyle\frac{g_{ii}(\vu)}{4(u^i+\nu)^2}&\rightarrow R^{ij}_{ij}=-1,\\
\displaystyle\frac{g_{ii}(\vu)}{8(u^i+\nu)^3}&\rightarrow
R^{ij}_{ij}=-2\nu-C^i(\vu)-C^j(\vu).
\end{split}
\end{equation}
The flatness of the pencil of metrics (\ref{pencil}) can be
obtained from the results in \cite{Du1}. An elegant proof of the
flatness of the metrics, valid for any genus can be obtained in a
more convenient set of coordinates $u^i\rightarrow
\displaystyle\frac{1}{u^i+\nu}$ that makes the spectral curve odd
with a branch point at infinity. We use these coordinates in the
next section to compare the Whitham equations for CH and for KdV.

\begin{remark}
The non-existence of a flat Egorov metric is related to the
non-existence of conservation laws of the form $a_t=b_x$ and
$b_t=c_x$ \cite{TP}. Furthermore the   non-existence of a flat
Egorov metric implies that the CH modulation equations cannot be
associated to a Frobenius manifold. We recall that, under certain
assumptions,  a  flat pencil of contravariant metrics on a
manifold induces a  Frobenius structure on it \cite{Du2}. One of
the assumptions is the requirement that one of the two flat
metrics is an Egorov metric. Therefore the geometric structure of
the CH modulation equations is substantially different from that
of the KdV modulation equations.
\end{remark}

To any flat diagonal Riemannian controvariant metric $g^{ii}$,
Dubrovin and Novikov \cite{DN} associate a local homogenous
Poisson bracket of hydrodynamic type
\[
\{F,G\}=\int\displaystyle\frac{\delta F}{\delta u^i}
A^{ij}\displaystyle\frac{\delta G}{\delta u^j}
\]
defined by the Hamiltonian operator
\begin{equation}
\label{flat}
A^{ij}=g^{ii}\delta^{ij}\displaystyle\frac{d}{dx}-g^{ii}\Gamma_{ik}^ju^k_x.
\end{equation}
Indeed in \cite{DN}, they prove that  $A^{ij}$ defines a  Poisson
tensor if and only if $g^{ii}$ is a flat non-degenerate  metric
and $\Gamma_{ik}^j$ are the Christoffel  symbols  of the
Levi-Civita connection compatible with the metric (the metric is
not necessary diagonal in their formulation). If the metric
$g^{ii}$ is not flat, the Poisson tensor needs to be modified
adding a non-local tail \cite{ferapontov},\cite{FM} of the form
\begin{equation}
\label{constantcurv}
A^{ij}=g^{ii}\delta^{ij}\displaystyle\frac{d}{dx}-g^{ii}\Gamma_{ik}^ju^k_x+c
u^i_x \left(\displaystyle\frac{d}{dx}\right)^{-1}u^j_x,
\end{equation}
for metrics of constant curvature $c$, and of the form
\begin{equation}
\label{conformal}
A^{ij}=g^{ii}\delta^{ij}\displaystyle\frac{d}{dx}-g^{ii}\Gamma_{ik}^ju^k_x+\eta^i
u^i_x \left(\displaystyle\frac{d}{dx}\right)^{-1}u^j_x+u^i_x
\left(\displaystyle\frac{d}{dx}\right)^{-1}\eta^ju^j_x,
\end{equation}
for conformally flat metrics. In the above relation the affinors
$\eta^j$ satisfy the equations
\[
R^{ij}_{ij}=\eta^i+\eta^j,
\]
where $R^{ij}_{ij}$ is the curvature tensor.

In the following let $g_{ii}(\vu)$ be as in (\ref{f1CH}). Let
$A^{ij}_1$ and $A^{ij}_2$ be the Hamiltonian operators of the form
(\ref{flat}) that correspond to the flat metric $g_{ii}(\vu)$ and
to $\displaystyle\frac{g_{ii}(\vu)}{2(u^i+\nu)}$, respectively.
The linear combination $A^{ij}_1+\lambda A^{ij}_2$ is an
Hamiltonian operator for any $\lambda$ because
$g_{ii}(\vu)+\lambda g_{ii}(\vu)/u^i$ is a flat pencil of metrics
for any $\lambda$.

\noindent
The Hamiltonian structure obtained in (\ref{modH}) coincides with
Hamiltonian operator $A_1$. Indeed the  coordinates
$k=k(u^1,u^2,u^3)$, $\beta=\beta(u^1,u^2,u^3)$ and
$J=J(u^1,u^2,u^3)$ defined in (\ref{modH}) are the densities of
the Casimirs for the Hamiltonian operator $A_1^{ij},$ namely
\[
A_1^{ij}\displaystyle\frac{\delta k}{\delta u^j}=A_1^{ij}
\displaystyle\frac{\delta \beta}{\delta
u^j}=A_1^{ij}\displaystyle\frac{\delta J}{\delta u^j}=0,
\]
and therefore they are the flat coordinates for the metric
$g_{ii}(\vu)$ \cite{DN}. Defining \[ h_0:=\displaystyle \oint
md\theta =\beta=-2\displaystyle\frac{P_2(-\nu)}{P_1(-\nu)}-\nu,\]
where $P_1$ and $P_2$ have been defined in (\ref{P1}) and
(\ref{P2})
 respectively, we obtain the
Hamiltonian densities $h_k=h_k(\vu)$, $k\geq 0$, by the recursion
relation
\[
A_2^{ij}\displaystyle\frac{\partial h_k}{\partial u^j}=A_1^{ij}
\displaystyle\frac{\partial h_{k+1}}{\partial u^j},\quad k\geq 0.
\]
The CH modulation hierarchy takes the bi-Hamiltonian form
\[
\partial_{t_k} u^i=A^{ij}_1\displaystyle \frac{\delta h_{k+1}}{\delta u^j} =
A^{ij}_2\displaystyle\frac{\delta h_k}{\delta u^j},\quad i=1,2,3.
\]
\begin{theorem}
For $k=1$ and $t_1=t$, the above equations coincide with the
modulation equations obtained in (\ref{whithamu}) and take the
bi-Hamiltonian form
\[
u_t^i={A}_1^{ij}\displaystyle\frac{\delta h_2}{\delta
u^j}={A}_2^{ij}\displaystyle\frac{\delta h_1}{\delta u^j}
\]
where the Hamiltonian densities are the averaged Hamiltonians
\[
h_2 = \frac{1}{2}\oint ({\tt u}^3+{\tt u}{\tt u}_x^2+2\nu {\tt u}^2)d\theta,
\quad h_1 = \frac{1}{2}\oint ({\tt u}^2+{\tt u}_x^2)d\theta.
\]
Moreover, the generating function for the Hamiltonian densities
$h_k$, $k\geq 0$, is given by the  coefficients of the expansion
as $\lambda \rightarrow\infty$ of the differential
\[
\displaystyle\frac{\Omega_{\nu}(\lambda)}{\left\{\res[\lambda=-\nu]
\displaystyle\frac{(\sigma_1(\lambda))^2}{d\lambda}\right\}^{\frac12}}\rightarrow
-(\xi_0+\xi_1 \displaystyle\frac{1}{\lambda}+\xi_2
\displaystyle\frac{1}{\lambda^2}+\dots)
\displaystyle\frac{d\lambda}{\lambda^2}.
\]
In particular, the first Hamiltonians (modulo Casimirs) are
\begin{equation}
\label{CHexpH}
 h_0=2\xi_0-\nu=\beta,\quad h_1=2\xi_1+2\nu\xi_0,\quad
h_2=\displaystyle\frac{8}{3}\xi_2+6 \nu\xi_1.
\end{equation}
\end{theorem}
In the following we write the CH-modulation equation in
Hamiltonian form with respect to a nonlocal Hamiltonian operator
of Mokhov-Ferapontov and Ferapontov type. Let $A_3^{ij}$ be the
Hamiltonian operator of the form (\ref{constantcurv}) associated
to the metric $\displaystyle\frac{g_{ii}(\vu)}{4(\nu+u^i)^2}$ of
constant curvature $c=1$, and let $A^{ij}_4$ be the Hamiltonian
operator of the form (\ref{conformal}) associated to the
conformally flat metric
$\displaystyle\frac{g_{ii}(\vu)}{8(\nu+u^i)^3}$ with affinors
$\eta^i= C^i+\nu$, $i=1,2,3$.

The CH modulation equations (\ref{whithamu}) can also
be written in a non-local Hamiltonian form
\[
\partial_{t} u^i =A^{ij}_3\displaystyle\frac{\delta h_{0}}{\delta u^j}=
A^{ij}_4\displaystyle\frac{\delta h_{-1}}{\delta u^j},\quad
i=1,2,3,
\]
where
\[
h_{-1}=1-\displaystyle\frac{\nu}{\left\{\res[\lambda=-\nu]\displaystyle\frac{(\sigma_1(
\lambda))^2}{d\lambda}\right\}^{\frac12}}.
\]
We remark that the Hamiltonian operators $A_3 $ and $A_4$ can be
obtained from the recursion $A_3=\mathcal{R}^2A_1$ and
$A_4=\mathcal{R}^3A_1 $ where $\mathcal{R}=A_2A_1^{-1}$ is the
recursion operator and $A_1^{-1}$ denotes the inverse of $A_1$.

In the limit when two Riemann invariants coalesce, the Hamiltonian
operators $A_1^{ij}$ and $A_2^{ij}$ reduce to the ``dispersionless
limit'' of the Poisson operators of the CH equation
\[
\mathcal{P}_{1}=-\displaystyle\frac{d}{dx},\quad
\mathcal{P}_{2}=-2({\tt u}+\nu)\displaystyle\frac{d}{dx}-{\tt u}_x,
\]
respectively. In the same limit, the Hamiltonian operators
$A_3^{ij}$ and $A_4^{ij}$ reduce to $\mathcal{P}_{2}
\mathcal{P}_{1}^{-1}\mathcal{P}_{2}$ and $\mathcal{P}_{2}
\mathcal{P}_{1}^{-1} \mathcal{P}_{2} \mathcal{P}_{1}^{-1}
\mathcal{P}_{2}$, respectively.

\subsection{Integration of the Whitham equations}
In this   subsection  we show how to integrate the Whitham
equations (\ref{whithamu}). All hydrodynamic systems  satisfying
(\ref{tsarev}) are integrable, via the generalized hodograph
transform introduced by Tsarev \cite{T}. Indeed such systems (not necessarily with
local Hamiltonian) possess an infinite number of commuting flows
\[
\partial_{t^\prime}u^i = w^i(\vu) \partial_x u^i,
\]
where the $w^i$ are solutions of the linear overdetermined system
\begin{equation}
\label{tsarev1} \frac{\partial_{u^j} w^i}{w^i-w^j} =
\frac{\partial_{u^j} C^i}{C^i-C^j}\, , \quad\quad i\neq j,
\end{equation}
where $C^{i}=C^{i}(u^1,u^2,u^3)$, $i=1,2,3,$ are the speeds in
(\ref{velu}). Then the solution
$\vu(x,t)=(u^1(x,t),u^2(x,t),u^3(x,t))$ of the so-called hodograph
transform
\begin{equation}
\label{ch0} x=-C^{i}(\vu)\,t+w^{i}(\vu) \,\quad i=1,2,3\,,
\end{equation}
satisfies the  system (\ref{whithamu}). Conversely, any solution
$(u^1(x,t),u^2(x,t),u^3(x,t))$  of (\ref{whithamu}) can be
obtained in this way. For monotone decreasing initial data
$x=f(u)|_{t=0}$, the general solution of the system
(\ref{tsarev1}) can be obtained following the work of Fei-Ran Tian
\cite{feiran} and the algebraic-geometric integration of Krichever
\cite{Krichever}. For simplicity, we restrict ourselves to the
case $\nu=0$.
\begin{proposition}
For $\nu=0$, and monotone increasing initial data
$x=f(u)|_{t=0}$, the solution of the system
(\ref{tsarev1}) is
\begin{equation}
w^i(\vu)=q(\vu)+\left(C^i(\vu)-u^1-u^2-u^3\right)
\displaystyle\frac{\partial q(\vu)}{\partial u^i},
\end{equation}
where the function $q=q(\vu)$ solves the linear over-determined
system of Euler-Poisson-Darboux type
\begin{equation}
\begin{split}
\label{euler}
&\partial_{u^i}q(\vu)-\partial_{u^j}q(\vu)
=2(u^i-u^j)\partial_{u^i}\partial_{u^j}q(\vu),\\
&q(u, u, u)=f(u).
\end{split}
\end{equation}
\end{proposition}
The proof of the above statement follows from \cite{feiran1}.
Equation (\ref{euler}) can be integrated and the explicit
expression of the function $q(u^1,u^2,u^3)$ is
\[
q(u^1,u^2,u^3)=\displaystyle\frac{1}{2\sqrt{2\pi}}\int_{-1}^1\int_{-1}^1
\displaystyle\frac{\displaystyle
f\left(\frac{1+\mu}{2}\frac{1+\eta}{2} u^1+
\frac{1+\mu}{2}\frac{1-\eta}{2}u^2+\frac{1-\mu}{2}u^3\right) }
{\sqrt{(1-\mu)(1-\eta^2)}}d\mu d \eta.
\]

\section{CH modulation  equations versus KdV and reciprocal
transformations}

In this section we compare CH and KdV  modulation equations. We
start recalling the reciprocal transformation which links the CH
equation to the first negative KdV flow. We show that the
 CH modulation equations (\ref{whithamu}) are transformed to the
modulation equations of the first negative KdV flow by the
averaged reciprocal transformation. Finally we compare the
averaged Hamiltonian operators of the two systems.

\subsection{A reciprocal transformation between Camassa Holm
equation and the first negative flow of KdV} In this subsection we
summarize the relation between the Camassa Holm equation and the
first negative flow of KdV hierarchy \cite{Fu}. The (associated)
Camassa-Holm equation is transformed into the first negative KdV
flow by a reciprocal transformation. In the following, to
distinguish between CH and KdV, we use $(x,t)$ for Camassa-Holm
variables and $(y,\tau_-)$ for the KdV variables.

The change of dependent variable $\rho^2=m+\nu$ transforms the
Camassa--Holm equation
\[
m_t = -2m{\tt u}_x-{\tt u}m_x-2\nu {\tt u}_x,\quad\quad m ={\tt u}-{\tt u}_{xx}.
\]
into the associated Camassa-Holm equation
\begin{equation}\label{CHp}
\left\{ \begin{array}{l} \rho_t =- \Big( {\tt u}\rho \Big)_x,\\
\rho^2 = {\tt u}-{\tt u}_{xx}+\nu,
\end{array}\right.
\end{equation}
which, via the reciprocal transformation introduced by
Fuchssteiner \cite{Fu},
\begin{equation}\label{TST}
\left\{ \begin{array}{l} dy = \rho dx-  {\tt u}\rho dt,\\
d\tau_- = dt,
\end{array}\right.
\end{equation}
is finally transformed into
\begin{equation}\label{CHRp}
\left\{ \begin{array}{l} {\tt u}= \displaystyle \rho^2-\nu -
\rho_{y\tau_-} +\frac{\rho_{\tau_-}
\rho_y}{\rho},\\
\displaystyle \left( \frac{1}{\rho}\right)_{\tau_-} = 2\rho\rho_y
-\left( \rho \Big(\log \rho\Big)_{y\tau_-} \right)_y.
\end{array}
\right.
\end{equation}
The transformation (\ref{TST}) is a reciprocal transformation because
the one-form $ \rho dx-  {\tt u}\rho dt$ is closed with respect to the CH flow.

The equation (\ref{CHRp}) is equivalent to the first negative flow
of the KdV hierarchy
\begin{equation}\label{kdvneg1}
\left( \partial_y^2 + 2U + U_y \partial_y^{-1} \right) U_{\tau_-}
=0,
\end{equation}
under the condition $U_{\tau_-} =- 2\rho_y$. Equation
(\ref{kdvneg1}) may be re-expressed as
\begin{equation}\label{kdvnegp}
\left\{ \begin{array}{l} \displaystyle U =\frac{\rho_y^2
-2\rho\rho_{yy} -1}{2\rho^2},\\
\displaystyle \left( \frac{\rho_y^2 -2\rho\rho_{yy}
-1}{4\rho^2}\right)_{\tau_-} = -\rho_y.
\end{array}
\right.
\end{equation}
Finally, we observe that $\displaystyle \int \rho(x,t) dx$ is a
Casimir of the second Hamiltonian operator of the Camassa--Holm
equation, $P_2 =m\partial_x +\partial_x m +2\nu\partial_x$.

The 2$\pi$-periodic solutions of the first negative KdV flow
(\ref{kdvneg1}), $U(\Theta)$, $\Theta=\mathcal{K}y-\Omega \tau$, satisfy
\begin{equation}
\label{solpU}
d\Theta = \frac{\mathcal{K} dU}{\sqrt{-U^3+\alpha U^2 -\beta U+\gamma}}.
\end{equation}
We may express such solutions also in the form $\rho(\Theta)$ or
${\tt u}(\Theta)$ and we easily get
\begin{equation}\label{solCHRp}
\left\{\begin{array}{l} \displaystyle {\tt u}_\Theta^2 =
\frac{1}{C^2\mathcal{K}^2} \left({\tt u}-c)({\tt u}^3-(c-2\nu){\tt u}^2+2B{\tt u}
-2A\right).
\\
\displaystyle \rho(\Theta) = -\frac{C}{{\tt u}(\Theta)-c},\\
\displaystyle U(\Theta)= -\frac{2}{{\tt u}(\Theta)-c},
\end{array}\right.
\end{equation}
where $C =\Omega/\mathcal{K}$ and $A,B,c$ are the constants defined in (\ref{periodic2})
which satisfy
\begin{equation}\label{alphaC}
Bc+\nu
c^2-A= C^2,\quad\displaystyle \alpha=-\frac{c^2+2B+4\nu c}{C^2}.
\end{equation}
The periodic solutions $u(\theta)$, $\rho(\theta)$, $\theta = k x
-\omega t$ of the (associated) Camassa--Holm equation, satisfy
\[
d\theta = \frac{k ({\tt u}-c)d{\tt u}}{\sqrt{({\tt u}-c) ({\tt
u}^3-(c-2\nu){\tt u}^2+2B{\tt u} -2A})}.
\]
We observe that the reciprocal transformation sends
2$\pi$-periodic solutions in $\Theta$ into 2$\pi$-periodic
solution in $\theta$ (and vice versa). Indeed, let $T$ be the
period of ${\tt u}(\theta)$, then
\[
T =\frac{k}{\mathcal{K}} \int_0^{2\pi} \frac{c -{\tt u}}{C}d\Theta
= k \oint \frac{ ({\tt u}-c)d{\tt u}}{\sqrt{({\tt u}-c)({\tt u}^3-
(c-2\nu){\tt u}^2+2B{\tt u} -2A})}=2\pi.
\]
It is then natural to expect that the average of the reciprocal
transformation connects the CH modulation equations to the modulation
equations of the first negative KdV flow.

\subsection{The modulation  equations of the negative KdV flow}\label{change}
In this subsection, we compute the modulation equations of the KdV
negative flow in the Riemann invariant coordinates used in the literature,
 namely the branch points  $\beta^1,\beta^2,\beta^3$, of the odd
elliptic curve \cite{W}
\begin{equation}
\label{wsurface} w^2=(\eta-\beta^1)(\eta-\beta^2)(\eta-\beta^3),
\end{equation}
where \[\beta^1+\beta^2+\beta^3=\alpha,\] with $\alpha$ defined in
(\ref{solpU}). It turns out from (\ref{alphaC}) that
\begin{equation}
\label{betau} \beta^i=\displaystyle\frac{1}{u^i+\nu}>0,\quad
i=1,2,3,
\end{equation}
where $u^i$ are the CH Riemann invariants defined in theorem~\ref{theoremRI}.
On the Riemann surface
(\ref{wsurface}) we define the second kind normalized differential
\begin{equation}
\label{dp}
dp(\eta)=\displaystyle\frac{\eta+\alpha_1}{\sqrt{(\eta-\beta^1)
(\eta-\beta^2)(\eta-\beta^3)}}d\eta,
\end{equation}
where  $\alpha_1$ is uniquely determined by the normalization condition
\[
\oint_{a}dp(\eta)=0.
\]
In the KdV literature $dp(\eta)$ is known as quasi-momentum. Now
we can  compute the velocities of the modulation equations of the negative KdV
flow
\begin{proposition}
The one phase Whitham equations of the first negative KdV flow are
\begin{equation}\label{whithkdvneg}
\partial_{\tau_-} \beta^i + v^i ({\bf \beta})\partial_y \beta^i =0,
\end{equation}
where
\begin{equation}
\label{velkdvneg}
v^i(\beta):= \frac{\partial_i \Omega(\beta)}{\partial_i \mathcal{K}(\beta)} =
\frac{2}{\sqrt{\beta^1\beta^2\beta^3}} \left(1-\frac{\prod_{j\not =i}
(\beta^i-\beta^j)}{\beta^i(\beta^i+\alpha_1)}\right),
\end{equation}
with $\alpha_1$ as in (\ref{dp}). In the above relations $\Omega$ and
$\mathcal{K}$ are the frequency and
wave-number of the one-phase KdV negative flow.
\end{proposition}

The proof of the proposition is as follows. From (\ref{solpU}) and
(\ref{solCHRp}), we immediately get $\displaystyle
\mathcal{K}=2\pi\mathcal{J}_0^{-1}$ and $\displaystyle\Omega= C\mathcal{K} =
4\pi\mathcal{J}_0^{-1} (\beta^1\beta^2\beta^3)^{-1/2}$, where
\[
\mathcal{J}_0 =\oint
\frac{d\lambda}{\sqrt{(\lambda-\beta^1)(\lambda-\beta^2)(\lambda-\beta^3)}}.
\]
Then the expressions
for the velocities (\ref{velkdvneg}) are computed from the
definition using the following variational formula
\begin{equation}\label{varj0}
\frac{\partial \mathcal{J}_0}{\partial \beta^i}=
\displaystyle\frac{1}{2}\mathcal{J}_0\frac{\beta^i+\alpha_1}{\prod_{j\not
=i} (\beta^i-\beta^j)},\quad i=1,2,3.
\end{equation}
\begin{remark}
In the limit $\beta^2=\beta^3$, the equations (\ref{whithkdvneg})
converge to $\displaystyle
\beta^1_{\tau_-}=-2(\beta^1)^{-\frac{3}{2}}\beta^1_{y}$, which is
the dispersionless limit of the first negative  KdV flow.
\end{remark}

Next, we write the negative KdV equations in Hamiltonian form. In
the $\beta$s' coordinates, the flat metric associated to the first
local KdV-Whitham Hamiltonian operator is \cite{Du1}
\begin{equation}
\label{metricKdV} g^{KdV}_{ii}(\boldsymbol{\beta})=
\res[\eta=\beta^i]\left\{\displaystyle\frac{(dp(\eta))^2}{d\eta}\right\},\quad
i=1,2,3,
\end{equation}
where the differential $dp$ has been defined in (\ref{dp}).
\begin{remark}
The flat metrics $g^{KdV}_{ii}(\boldsymbol{\beta})$ and
$g^{KdV}_{ii}(\boldsymbol{\beta})/\beta^i$ can be related to a
Frobenius manifold defined on the moduli space of elliptic curves
$w^2=(\eta-\beta^1)(\eta-\beta^2)(\eta-\beta^3)$ \cite{Du2}.
\end{remark}
Let  $J_1$ and $J_2$ be  the local KdV Hamiltonian operators of
the form (\ref{flat}) associated to the flat  metrics
$\displaystyle
\displaystyle\frac{1}{8}g^{KdV}_{ii}(\boldsymbol{\beta})$ and
$\displaystyle\displaystyle\frac{g^{KdV}_{ii}(\boldsymbol{\beta})}{4\beta^i}$.
Let  $J_3$ be  the nonlocal Hamiltonian operator of the form
(\ref{constantcurv}) associated to the metric $\displaystyle
\displaystyle\frac{g^{KdV}_{ii} ( \boldsymbol{\beta})}{
2(\beta^i)^2}$ of constant curvature
$R^{ij}_{ij}=-\displaystyle\frac{1}{2}$. Let $J_4$ be the nonlocal
Hamiltonian operator of the form (\ref{conformal}) associated to
the conformally flat metric $\displaystyle
\displaystyle\frac{g^{KdV}_{ii}(\boldsymbol{\beta})}{(\beta^i)^3}$
with Riemannian curvature $\displaystyle
R_{ij}^{ij}=-\displaystyle\frac{1}{8}(w^i_++w^j_+),$ where
\begin{equation}
\label{w+}
w^i_+=  \left( \beta^1+\beta^2+\beta^3 +\frac{2\prod_{j\not =i}
(\beta^i-\beta^j)}{\beta^i+\alpha_1}\right),
\end{equation}
with $\alpha_1$  defined in (\ref{dp}). The $w^i_+$,
$i=1,\dots,3$, are the velocities (originally obtained by Whitham
\cite{W}) of the usual positive KdV modulated flow
\[
\displaystyle\frac{\partial \beta^i}{\partial
\tau_+}+w_+^i(\boldsymbol{\beta}) \displaystyle\frac{\partial
\beta^i }{\partial y}=0.
\]
\begin{remark}\label{46}
The Hamiltonian operator $J_1$ corresponds to the average  over a
one-dimensional torus of the Gardner-Zakharov KdV Hamiltonian
structure $\mathcal{P}_1=8\partial_y$ while the Hamiltonian
operator $J_2$ corresponds to the average  over a one-dimensional
torus of the Lenard-Magri local Hamiltonian structure
$\mathcal{P}_2=2\partial_{yyy}+2U\partial_y+2\partial_y U$. We use
this unusual normalization in order to be consistent with the
normalization of the CH equation. Defining the recursive operator
$\mathcal{R}=\mathcal{P}_2(\mathcal{P}_1)^{-1}$, we obtain the
family of non-local Hamiltonian operators
$\mathcal{P}_{k+1}=\mathcal{R}\mathcal{P}_{k},$ $k\geq 1$. The
averaged KdV non-local Hamiltonian operators $J_3$ and $J_4$
corresponds to the average  over a one-dimensional torus of the
non-local Hamiltonian operators $\mathcal{P}_{3}$ and
$\mathcal{P}_{4}$, respectively.
\end{remark}
The modulation equations of the negative KdV  flow
can be written in Hamiltonian form with Hamiltonian operator $J_1$ and
Hamiltonian density ${\cal H}_0$, which
is the average over the one dimensional torus of the Casimir
generating the KdV negative flow, namely
\begin{equation}\label{K0}
{\cal H}_0 = \displaystyle\oint \frac{d\Theta}{\rho (\Theta)}.
\end{equation}
\begin{lemma}\cite{Du2}
In the $\beta$s' coordinates
\[
{\cal H}_0=ip(0),
\]
where  $p(\eta)$ is the Abelian integral of the quasi-momentum
$dp(\eta)$ defined in (\ref{dp}).
\end{lemma}
To prove the lemma, we compute the integral in (\ref{K0})
in the $\beta$s' coordinates obtaining
\[ {\cal H}_0=\displaystyle\oint
\frac{d\Theta}{\rho(\Theta)}
=-\sqrt{\beta^1\beta^2\beta^3}\alpha_0=
-i\left.\displaystyle\frac{(\Lambda_0(\eta))}{2d\xi}\right|_{\xi=0},\quad
\eta=\displaystyle\frac{1}{\xi^2},
\]
where
\begin{equation}\label{Lambda00}
\Lambda_0(\eta)=
\sqrt{-\beta^1\beta^2\beta^3}\displaystyle\frac{\displaystyle\frac{1}{\eta}
+\alpha_0}{\sqrt{(\eta-\beta^1)
(\eta-\beta^2)(\eta-\beta^3)}}d\eta.
\end{equation}
$\Lambda_0(\eta)$ is a normalized third kind differential with
first order poles at the points  $O^{\pm}=(0,\pm
\sqrt{-\beta^1\beta^2\beta^3})$ with residue $\pm 1 $,
respectively. The constant $\alpha_0$ in (\ref{Lambda00}) is
uniquely determined by the condition $\displaystyle
\oint_{a}\Lambda_0(\eta)=0.$ From the Riemann bilinear relation,
it is finally immediate to verify that
\begin{equation}\label{aiuto}
\mathcal{H}_0=-i
\left.\displaystyle\frac{(\Lambda_0(\eta))}{2d\xi}\right|_{\xi=0}=ip(0),\quad
\eta=\displaystyle\frac{1}{\xi^2},
\end{equation}
where $p(\eta)$ is the Abelian integral of the quasi-momentum
$dp(\eta)$ defined in (\ref{dp}).
Finally the following
result holds.
\begin{lemma}
The first negative KdV averaged flow (\ref{whithkdvneg}) can be written in the
Hamiltonian form
\begin{equation}\label{hamkdv}
\beta^i_{\tau_-}=J^{ij}_1\displaystyle\frac{\delta
\mathcal{H}_{0}}{\delta\beta^j}= J^{ij}_2\displaystyle\frac{\delta
\mathcal{H}_{-1}}{\delta\beta^j}=J^{ij}_3
\displaystyle\frac{\delta
\mathcal{H}_{-2}}{\delta\beta^j}=J^{ij}_4\displaystyle\frac{\delta
\mathcal{H}_{-3}}{\delta\beta^j},
\end{equation}
where $\mathcal{H}_{0}$ is the Casimir of $J_2$ defined in
(\ref{K0}) and the Hamiltonian densities are determined by the
recursion scheme
\[
J^{ij}_1\displaystyle\frac{\delta \mathcal{H}_{-s}}{\delta\beta^j}
= J^{ij}_2\displaystyle\frac{\delta
\mathcal{H}_{-s-1}}{\delta\beta^j},\quad s\geq 0.
\]
\end{lemma}
The Hamiltonian  $\mathcal{H}_{-s}$, $s\geq 1$, are generated by
the expansion for $\eta \rightarrow 0$ of the quasi-momentum
$dp(\eta)$. Indeed,
\begin{equation}
\label{expkdv0} i \int_{(\eta,-w)}^{(\eta,w)}dp(\xi)=
\displaystyle\frac{1}{2}(\mathcal{H}_0+\eta\mathcal{H}_{-1}
+\eta^2\mathcal{H}_{-2}+\displaystyle\frac{3}{4}\eta^3
\mathcal{H}_{-3}+\dots),\quad \eta\rightarrow 0.
\end{equation}

\subsection{CH versus KdV modulation equations\label{sec34}}

In the previous section we computed the modulation equations of the first negative
KdV flow. In this subsection, we compute the average of the
reciprocal transformation defined in section 4.1 and we show that
the negative KdV modulation equations are transformed to the CH
modulation equations. Finally, both KdV-Whitham and CH-Whitham
systems are Hamiltonian systems, so we end the section
investigating how the reciprocal transformation acts on the
Hamiltonian  structures of the two systems.

To compare KdV and CH, we first need to reduce the even spectral
curve of CH to the odd spectral curve of KdV. The natural change
of coordinates $(\lambda,y)\rightarrow (\eta,w)$
\[
\eta=\displaystyle\frac{1}{\lambda+\nu},\quad
w^2=-\displaystyle\frac{y^2}{(\lambda+\nu)^4\prod_{i=1}^3
(\nu+u^i)},
\]
maps the even spectral curve
$y^2=(\lambda+\nu)(\lambda-u^1)(\lambda-u^2)(\lambda-u^3)$ to the
odd KdV spectral curve
\[
w^2=(\eta-\beta^i)(\eta-\beta^2)(\eta-\beta^3),\quad
\beta^i=\displaystyle\frac{1}{\nu+u^i}.
\]
The differentials $\Omega_{\nu}(\lambda)$ and $\sigma_1(\lambda)$
defined in (\ref{omeganu}) and (\ref{P1}) transform to
\begin{equation}
\label{tem0} \Omega_{\nu}(\lambda)\rightarrow
\displaystyle\frac{1}{2}dp(\eta),
\end{equation}
\begin{equation}
\label{Lambda0} \sigma_1(\lambda)\rightarrow \Lambda_0(\eta),
\end{equation}
with $dp(\eta)$ as in (\ref{dp}) and $\Lambda_0(\eta)$ as in
(\ref{Lambda00}). It follows from (\ref{tem0}) and (\ref{Lambda0})
that the change of coordinates $\beta^i=1/(\nu+u^i)$ transforms
the speeds $C^i(\vu)$ defined in (\ref{Ci}) to
\begin{equation}
\label{tildeCi} \tilde{C}^i(\boldsymbol{\beta})=
\displaystyle\frac{1}{\beta^1}+\displaystyle\frac{1}{\beta^2}+
\displaystyle\frac{1}{\beta^3}-\nu
+2 \frac{\alpha_{0}\prod_{j\neq i,j=1}^3(\beta^i-\beta^j)}{\beta^i
(\beta^i+\alpha_1)}.
\end{equation}

Now we show that the averaged reciprocal transformation maps the
modulation equations of the KdV negative flow to the CH modulation equations. Indeed
averaging over a period the inverse of (\ref{TST})
\[
dx = 1/\rho dy+ {\tt  u} d\tau_-,\quad\quad dt = d\tau_-,
\]
we get the averaged reciprocal transformation
\begin{equation}\label{art0}
dx =  dy\oint \frac{d\Theta}{\rho(\Theta)} +d\tau_-\oint
{\tt u}(\Theta)\displaystyle d\Theta\quad\quad dt=d\tau_-.
\end{equation}
\begin{proposition}
The averaged reciprocal transformation (\ref{art0}) takes the form
\begin{equation}
\label{art} dx = {\cal H}_0 dy +\mathcal{N} d\tau_-,\quad\quad
dt=d\tau_-,
\end{equation}
where ${\cal H}_0$ is the Casimir defined in (\ref{K0})
and
\begin{equation}
\label{M} {\cal N}= \frac{1}{\beta^1}+
\frac{1}{\beta^2}+\frac{1}{\beta^3}-\nu+2\alpha_0 =\frac{1}{2}(
\nabla {\cal H}_0 )^2-\nu,
\end{equation}
where $( \nabla {\cal H}_0
)^2=\sum_i(g^{KdV}_{ii})^{-1}(\partial_{\beta^i}\mathcal{H}_0)^2
$. Finally (\ref{art}) is a reciprocal transformation for the
KdV-Whitham negative flow.
\end{proposition}
To prove the proposition we observed that
the one form (\ref{art}) is closed by (\ref{M}). The proof of
(\ref{M}) follows from the identity $\displaystyle {\cal H}_0
=i p(0)$, and the variational formula $\displaystyle
\partial_{\beta^i}p(0)=\displaystyle\frac{1}{4}dp(\beta^i)\Lambda_0(\beta^i)$.

Next we show that the  modulation equations of the first negative KdV flow are mapped
by the reciprocal transformation (\ref{art}) to the CH modulation
equations.

\begin{proposition}\label{proprec}
The reciprocal transformation $dx = {\cal H}_0 dy +{\cal
N}d\tau_-$, $dt=d\tau_-$, where ${\cal H}_0$ and ${\cal N}$ are as
in (\ref{K0}) and (\ref{M}) respectively, transforms the
modulation equations (\ref{whithkdvneg}) of the first negative KdV
flow
\begin{equation}\label{KdV1}
\partial_{\tau_-}\beta^i + v^i ({\vb})\partial_{y} \beta^i =0,
\end{equation}
where $v^i(\vb)$, $i=1,\dots,3$ are as in (\ref{velkdvneg}) to the
CH modulation equations
\begin{equation}\label{CH1}
\partial_t \beta^i + {\tilde C}^i (\vb)\partial_x \beta^i =0,
\end{equation}
where the CH velocities are defined in (\ref{tildeCi}).
Viceversa, the inverse reciprocal transformation $\displaystyle
dy=\frac{1}{{\cal H}_0}dx-\frac{{\cal N}}{{\cal H}_0}dt$
transforms (\ref{CH1}) into (\ref{KdV1}).
\end{proposition}

Indeed plugging (\ref{art}) into (\ref{KdV1}) we get $\displaystyle
\partial_t \beta^i +V^i(\vb)
\partial_x \beta^i=0$, where
\[
V^i(\vb) =v^i(\vb) {\cal H}_0 +{\cal N}=\tilde{C}^i(\vb).
\]

\medskip

To compare the CH and KdV Hamiltonian structures, we express the
CH-Whitham Hamiltonian operators introduced in subsection 3.1 in
the $\beta$s' coordinates. In these coordinates, using
(\ref{aiuto}), (\ref{expkdv0}) and (\ref{tem0}), the CH flat
metric $g_{ii}(\vu)$ defined in (\ref{f1CH}) takes the form
\begin{equation}
\label{metricbeta} \tilde{g}_{ii}(\boldsymbol{\beta})=
\displaystyle\frac{-1}{(\beta^i)^3}\displaystyle\frac{\res[\eta=\beta^i]
\left\{\displaystyle\frac{(dp(\eta))^2}{d\eta}\right\}} {p(0)^2/4}
=
\displaystyle\frac{g^{KdV}_{ii}(\boldsymbol{\beta})}{(\beta^i)^3{\cal
H}_0^2}, \quad i=1,2,3.
\end{equation}
The other metrics defined in (\ref{f2CH}) transform, respectively,
to
\begin{equation}
\label{metricbeta1}
 \displaystyle\frac{g^{KdV}_{ii}( \boldsymbol{\beta})}{2(\beta^i)^2
{\cal H}_0^2}, \quad
\displaystyle\frac{g^{KdV}_{ii}(\boldsymbol{\beta})}{4\beta^i{\cal
H}_0^2},\quad \displaystyle\frac{
g^{KdV}_{ii}(\boldsymbol{\beta})}{8{\cal H}_0^2},
\end{equation}
and are, respectively, flat, of constant curvature and conformally
flat.

Let $\tilde{A}_1$ and $\tilde{A}_2$ be the local Hamiltonian
operators associated to the CH flat metrics $\displaystyle
\tilde{g}_{ii}(\boldsymbol{\beta})$ and $\displaystyle
\displaystyle\frac{1}{2} \tilde{g}_{ii}(\boldsymbol{\beta})
\beta^i$, respectively. Moreover, let $\tilde{A}_3$ and
$\tilde{A}_4$ be the non-local Hamiltonian operators associated to
the CH constant curvature and conformally flat metrics,
$\displaystyle \displaystyle\frac{1}{4}
\tilde{g}_{ii}(\boldsymbol{\beta}) (\beta^i)^2$ and $\displaystyle
\displaystyle\frac{1}{8}\tilde{g}_{ii}(\boldsymbol{\beta})(\beta^i)^3$,
respectively. Then, in the $\beta$s' coordinates the CH modulation
equations in Hamiltonian form are
\begin{equation}\label{hamch} \beta_t^i=\tilde{A}_1^{ij}\displaystyle\frac{\delta
\tilde{h}_2}{\delta
\beta^j}=\tilde{A}_2^{ij}\displaystyle\frac{\delta
\tilde{h}_1}{\delta
\beta^j}=\tilde{A}_3^{ij}\displaystyle\frac{\delta
\tilde{h}_0}{\delta
\beta^j}=\tilde{A}_4^{ij}\displaystyle\frac{\delta
\tilde{h}_{-1}}{\delta \beta^j}
\end{equation}
where the Hamiltonians $\tilde{h}_j$, $i=-1,\dots,2,$ are the
averaged conservation laws introduced in subsection 3.1 expressed
in the $\beta$s' coordinates. Indeed,  following (\ref{expkdv0}),
the positive CH Hamiltonians are obtained from the coefficients of
the expansion for $\eta\rightarrow 0$ of the differential
$\displaystyle\frac{dp(\eta)}{{\cal H}_0}$ and take the form
\begin{equation}
\label{Hbeta} \tilde{h}_{-1}=1-\displaystyle\frac{\nu}{\mathcal{
H}_0},\quad
\tilde{h}_0=\displaystyle\frac{\mathcal{H}_{-1}}{\mathcal{H}_0}-\nu,\quad
\tilde{h}_j=\displaystyle\frac{\mathcal{H}_{-j-1}}{\mathcal{H}_0}-\nu\displaystyle\frac{
\mathcal{H}_{-j}}{\mathcal{H}_0},\quad j\ge 1,
\end{equation}
where the $\mathcal{H}_{-k}$ are defined in (\ref{expkdv0}).
The following theorem by Ferapontov and Pavlov describes the
action of a reciprocal transformation for an Hamiltonian
hydrodynamic equation and we use it to clarify the relation
between the Hamiltonian structures of the averaged KdV and CH
Hamiltonian structures.

\begin{theorem}\label{ferap}\cite{FP}
Let $\displaystyle \beta^i_{\tau} = J^{ij} \frac{\partial
h}{\partial\beta^j}$ be an Hamiltonian system associated to the
local operator $\displaystyle \; J^{ij} =
g^{ii}\delta^{j}_{i}\frac{d}{dy} - g^{ii} \Gamma^j_{ik}
\beta_{y}^k$. Then, under the action of the reciprocal
transformation $\; dx=Ady + Bd\tau, \;dt=d\tau$, where $d(Ady +
Bd\tau )=0$, the transformed system is Hamiltonian $\displaystyle
\beta^i_t={\tilde J}^{ij}\frac{\partial {\tilde h}}{\partial
\beta^j},$ with nonlocal operator
\[
{\tilde J}^{ij} ={\tilde g}^{ii}\delta^{ij}\frac{d}{dx} -{\tilde
g}^{ii} {\tilde \Gamma}^j_{ik} \beta_x^k +{\tilde
w}^i\beta^i_x\left(\frac{d}{dx} \right)^{-1}\beta^j_x +\beta^i_x
\left( \frac{d}{dx}\right)^{-1} {\tilde w}^j \beta^j_x,
\]
and hamiltonian density ${\tilde h}=h/A$. The transformed metric
is ${\tilde g}_{ii} = g_{ii}/A^2$, ${\tilde \Gamma}$ is the
Levi--Civita connection and
\begin{equation}\label{curvcoe}
{\tilde w}^i= \nabla^i\nabla_i A \cdot A - \frac{1}{2} (\nabla
A)^2 = g^{ii} \left( \partial_i^2 A - \sum_j \Gamma^{j}_{ii}
\partial_j A \right) A -\frac{1}{2}\sum_j g^{jj} (\partial_j A)^2.
\end{equation}
Moreover, the transformed  metric is conformally flat with
curvature tensor
\[
{\tilde R}^{ij}_{ij} = {\tilde w}^i  +{\tilde w}^j.
\]
\end{theorem}

By the above theorem and by inspection of (\ref{hamkdv}) and
(\ref{hamch}), where the metrics have been defined in
(\ref{metricbeta}), (\ref{metricbeta1}) and (\ref{metricKdV}), we
conclude  that the local KdV-Whitham Poisson operators, $J_1$ and
$J_2$, are mapped to the nonlocal CH-Whitham Poisson operators
${\tilde A}_{4}$ and ${\tilde A}_{3}$, respectively, by the
reciprocal transformation $\displaystyle dx = {\cal H}_0 dy +{\cal
N}d\tau_-$, $dt=d\tau_-$.

\begin{remark}
The application of theorem~\ref{ferap} deserves a special
attention in the case $\nu \not =0$ for the  computation of the
Hamiltonian densities $\tilde{h}_k$. Indeed neither the
KdV--Hamiltonian operators $J_i$  nor the averaged negative KdV
Hamiltonian densities ${\cal H}_{-s}$, $s=0,1,2,3$, depend on the
parameter $\nu$ while the CH averaged Hamiltonian densities
${\tilde h}_{s}$, $s=-1,0,1,2,$ in (\ref{Hbeta}) do. To solve this
apparent contradiction, we recall that the Hamiltonian densities
are defined modulo Casimirs.

In particular, the Casimir $-\nu{\cal H}_0$ of the second
KdV-Whitham Hamiltonian operator is transformed to the CH
Hamiltonian density $-\nu$ which generates the constant flow term
associated to ${\tilde A}_3$. Similarly, the term $-\nu$ is mapped
by the reciprocal transformation to the CH Hamiltonian density
$-\nu/{\cal H}_0$ which generates the constant flow term
associated to ${\tilde A}_4$.
\end{remark}

Applying theorem \ref{ferap} to the local CH--Whitham Poisson
operators, we prove that the local Hamiltonian CH operators
${\tilde A}_{1}$ and ${\tilde A}_{2}$, are mapped to the nonlocal
KdV--Whitham Poisson operators $J_4$ and $J_3$, respectively, by
the reciprocal transformation $\displaystyle dy = 1/{\cal H}_0 dx
-{\cal N}/{\cal H}_0 dt$. By the same theorem \ref{ferap}, the
corresponding KdV Hamiltonian densities are related to the CH ones
by
\[
\tilde{h}_2\mathcal{ H}_0=\mathcal{H}_{-3},\quad
\tilde{h}_1\mathcal{H}_0=\mathcal{H}_{-2},
\]
where the above identities again hold modulo Casimirs of the
corresponding KdV Hamiltonian operators. We summarize the results
of this section in the following.

\begin{theorem}
The reciprocal transformation  $dx={\cal
H}_0dy+{\cal N}d\tau_-$, $dt=d\tau_-$,
  maps the KdV local Hamiltonian operators $J_1$, $J_2$ and the corresponding
Hamiltonian densities $\mathcal{H}_0$,  $\mathcal{H}_{-1}$ to
\[
J^{ik}_1\displaystyle\frac{\delta \mathcal{H}_0}{\delta
\beta^k}\longrightarrow \tilde{A}_4^{ik} \displaystyle\frac{\delta
{\tilde h}_{-1}}{\delta \beta^k}\
\]
\[
J^{ik}_2\displaystyle\frac{\delta \mathcal{H}_{-1}}{\delta
\beta^k}\longrightarrow \tilde{A}_3^{ik} \displaystyle\frac{\delta
{\tilde h}_{0}}{\delta \beta^k}\
\]
where ${\tilde A}_4$ is   the CH nonlocal Hamiltonian operator
associated to the covariant conformally flat metric $\displaystyle
\displaystyle\frac{g^{KdV}_{ii}(\boldsymbol{\beta})}{8\mathcal{H}_0^2}$
 and   ${\tilde h}_{-1}=1-\nu/\mathcal{ H}_0$
the corresponding   CH Hamiltonian density ( ${\tilde h}_{-1}
\mathcal{H}_0= \mathcal{H}_0$ modulo Casimir of $J_1$). The
Hamiltonian operator ${\tilde A}_3$ is   the CH nonlocal
Hamiltonian operator   associated to the constant curvature metric
$\displaystyle
\displaystyle\frac{g^{KdV}_{ii}(\boldsymbol{\beta})}{4\mathcal{H}_0^2\beta^i}$
 and
 ${\tilde h}_0=-\nu+\displaystyle\frac{\mathcal{H}_{-1}}{\mathcal{ H}_0}$ the
corresponding CH Hamiltonian density (
${\tilde h}_0{\cal H}_0={\cal H}_{-1}$ modulo a Casimir of $J_2$).
\noindent

\noindent The reciprocal transformation  $dy= 1/{\cal H}_0 dx
-{\cal N}/{\cal H}_0 dt$, $d\tau_-=dt$  maps the local CH
Hamiltonian operators ${\tilde A}_1$,  ${\tilde A}_2$, and
corresponding Hamiltonian densities
$\tilde{h}_2=\displaystyle\frac{\mathcal{H}_{-3}}{\mathcal{H}_0}-
\nu\displaystyle\frac{\mathcal{H}_{-2}}{\mathcal{H}_0}$,
$\tilde{h}_1=\displaystyle\frac{\mathcal{H}_{-2}}{\mathcal{H}_0}-
\nu\displaystyle\frac{\mathcal{H}_{-1}}{\mathcal{H}_0}$ to
\[
\tilde{A}_1^{ik} \displaystyle\frac{\delta {\tilde h}_{2}}{\delta
\beta^k}\longrightarrow J^{ik}_4\displaystyle\frac{\delta
\mathcal{H}_{-3}}{\delta \beta^k}\
\]
\[
\tilde{A}_2^{ik} \displaystyle\frac{\delta {\tilde h}_{1}}{\delta
\beta^k}\longrightarrow J^{ik}_3\displaystyle\frac{\delta
\mathcal{H}_{-2}}{\delta \beta^k}\
\]
where  the nonlocal KdV Hamiltonian operator $J_4$ is associated to
the covariant conformally flat metric $\displaystyle
\displaystyle\frac{g^{KdV}_{ii}(\boldsymbol{\beta})}{(\beta^i)^3}$
and $\tilde{h}_2\mathcal{H}_{0}=\mathcal{H}_{-3}$ modulo a Casimir
of $J_4$. The nonlocal KdV Hamiltonian operator $J_3$ is associated to
the covariant constant curvature metric $\displaystyle
\displaystyle\frac{g^{KdV}_{ii} ( \boldsymbol{\beta})}{
2(\beta^i)^2}$ and $\tilde{h}_1\mathcal{H}_{0}= \mathcal{H}_{-2}$
modulo a Casimir of $J_3$.

\end{theorem}

We conclude this section by illustrating with a table all the
reciprocal transformations between the various Hamiltonian
operators.

\begin{table}
\begin{tabular}{|c|c|c|c|c|}
    \hline
KdV
Poisson&$\mathcal{P}_1$&$\mathcal{P}_2$&$\mathcal{R}^2\mathcal{P}_1$&
$\mathcal{R}^3\mathcal{P}_1$\\
tensor&&&&\\
\hline
averaged KdV&$J_1$&$J_2$&$J_3$&$J_4$\\
Poisson tensor&&&&\\
\hline KdV
metric&$\displaystyle\frac{1}{8}g^{KdV}_{ii}(\boldsymbol{\beta})$&
$\displaystyle\frac{g^{KdV}_{ii}(\boldsymbol{\beta})}{4\beta^i}$&
$\displaystyle\frac{g^{KdV}_{ii}(\boldsymbol{\beta})}{2(\beta^i)^2}$&
$\displaystyle\frac{g^{KdV}_{ii}(\boldsymbol{\beta})}{(\beta^i)^3}$\\
&&&&\\
 Curvature tensor $R^{ij}_{ij}$ &$0 $
&$0$&$-1/2$&
$-(w^i_++w^j_+)/8$\\
\hline averaged KdV&$\mathcal{H}_0-\nu$
&$\mathcal{H}_{-1}-\nu\mathcal{H}_0$&$\mathcal{H}_{-2}-\nu\mathcal{H}_{-1}$&
$\mathcal{H}_{-3}-\nu\mathcal{H}_{-2}$\\
Hamiltonian&&&&\\
\hline Reciprocal&\multicolumn{2}{c}{$dx=\mathcal{H}_0 dy
+\mathcal{N}d\tau$}&\multicolumn{2}{|c|}{$\uparrow$}\\
transformation&\multicolumn{2}{|c|}{$\downarrow$}&
\multicolumn{2}{|c|}{$dy=\displaystyle\frac{1}{\mathcal{H}_0}dx
-\displaystyle\frac{\mathcal{N}}{\mathcal{H}_0}dt$}\\
 \hline
averaged CH&$\tilde{\mathcal{A}}_4$&$\tilde{\mathcal{A}}_3$&
$\tilde{\mathcal{A}}_2$&$\tilde{\mathcal{A}}_1$\\
Poisson tensor&&&&\\
\hline CH
metric&$\displaystyle\frac{1}{8\mathcal{H}_0^2}g^{KdV}_{ii}(\boldsymbol{\beta})$&
$\displaystyle\frac{g^{KdV}_{ii}(\boldsymbol{\beta})}{4\mathcal{H}_0^2\beta^i}$&
$\displaystyle\frac{g^{KdV}_{ii}(\boldsymbol{\beta})}{2\mathcal{H}_0^2(\beta^i)^2}$&
$\displaystyle\frac{g^{KdV}_{ii}(\boldsymbol{\beta})}{\mathcal{H}_0^2(\beta^i)^3}$\\
&&&&\\
Curvature tensor $R^{ij}_{ij}$ &$-2\nu-\tilde{C}^i-\tilde{C}^j $
&$-1$&$0$&
$0$\\
\hline averagedCH &$1-\displaystyle\frac{\nu}{\mathcal{H}_0}$
&$\displaystyle\frac{\mathcal{H}_{-1}}{\mathcal{H}_0}-\nu$&
$\displaystyle\frac{\mathcal{H}_{-2}}{\mathcal{H}_0}-\nu
\displaystyle\frac{\mathcal{H}_{-1}}{\mathcal{H}_0}$&
$\displaystyle\frac{\mathcal{H}_{-3}}{\mathcal{H}_0}-
\nu\displaystyle\frac{\mathcal{H}_{-2}}{{\mathcal{H}_0}}$\\
Hamiltonian&&&&\\
\hline
\end{tabular}
\caption{In the above table   $\mathcal{P}_1$, $\mathcal{P}_2$ and
$\mathcal{R}$ have been defined in remark~\ref{46}. The averaged
KdV Hamiltonian operators $J_i$, $i=1,\dots,4$ and corresponding metrics
have been defined at the begining of section~\ref{sec34}. The
quantities $\mathcal{H}_{-s}$ are defined in (\ref{K0}) and
(\ref{expkdv0}). The averaged KdV Hamiltonian densities for the
operators $J_i$, $i=1,2,3,4$ are defined modulo Casimirs. The CH
Hamiltonian operators $\tilde{A}_{s}$ are defined below formula
(\ref{metricbeta1}) and the corresponding Hamiltonians are defined
in (\ref{Hbeta}).} The speeds $w^i_+$ are defined in (\ref{w+}) and the
speeds $\tilde{C}^i$ are defined in (\ref{tildeCi}).
\end{table}
\vskip 1cm
{\bf Acknoledgments} We express our gratitude to A. Maltsev for
the many useful discussions and for pointing out to us the
Lagrangian averaging method for the Camassa-Holm equations. We
also thank B. Dubrovin, G. Falqui and D. Holm for the useful
discussions and comments of the manuscript. This work has been
partially supported by the GNFM-INdAM  Project "Onde nonlineari,
struttura tau e geometria delle variet\`a invarianti: il caso
della gerarchia di Camassa-Holm" and by the European Science
Foundation Programme MISGAM (Method of Integrable Systems,
Geometry and Applied Mathematics).

\end{document}